# The Groupon Effect on Yelp Ratings: A Root Cause Analysis

JOHN W. BYERS, Computer Science Department, Boston University
MICHAEL MITZENMACHER, School of Engineering & Applied Sciences, Harvard University
GEORGIOS ZERVAS, Computer Science Department, Yale University

Daily deals sites such as Groupon offer deeply discounted goods and services to tens of millions of customers through geographically targeted daily e-mail marketing campaigns. In our prior work we observed that a negative side effect for merchants using Groupons is that, on average, their Yelp ratings decline significantly. However, this previous work was essentially observational, rather than explanatory. In this work, we rigorously consider and evaluate various hypotheses about underlying consumer and merchant behavior in order to understand this phenomenon, which we dub the Groupon effect. We use statistical analysis and mathematical modeling, leveraging a dataset we collected spanning tens of thousands of daily deals and over 7 million Yelp reviews. In particular, we investigate hypotheses such as whether Groupon subscribers are more critical than their peers, or whether some fraction of Groupon merchants provide significantly worse service to customers using Groupons. We suggest an additional novel hypothesis: reviews from Groupon subscribers are lower on average because such reviews correspond to real, unbiased customers, while the body of reviews on Yelp contain some fraction of reviews from biased or even potentially fake sources. Although we focus on a specific question, our work provides broad insights into both consumer and merchant behavior within the daily deals marketplace.

Categories and Subject Descriptors: J.4 [**Social and Behavioral Sciences**]: Economics

## 1. INTRODUCTION

Groupon is the largest of the daily deals businesses: each day, in a number of markets (generally cities), they offer Groupons, which are deals providing a coupon for some product or service at a substantial discount, generally approximately 50% off the list price. Daily deal sites such as Groupon represent a novel approach to Internet marketing that tap into local markets, and based on the massive scale and rapid growth of such sites, the business model is worthy of further examination.

In previous work examining Groupon, we measured and evaluated aspects of Groupon's operational strategy, as well as observed the impact of customer behaviors including word-of-mouth effects and how running a Groupon offer affects a merchant's reputation [Byers et al. 2012]. One specific finding that received significant

---

This work is supported by the National Science Foundation.
Author's addresses: J. W. Byers, Computer Science Department, Boston University; M. Mitzenmacher, School of Engineering & Applied Sciences, Harvard University; G. Zervas, Computer Science Department, Yale University.

interest and attention was that Yelp reviews that contained the word "Groupon" provided, on average, significantly lower ratings than reviews that did not, to the extent that it significantly lowered the average rating for businesses that used Groupon. In the original paper, we did not examine reasons to explain this finding, leaving it to future work. Our analysis met with a variety of reactions, ranging from disbelief, to a number of plausible explanations for why this phenomenon should be expected: for example, Groupon users are fussy reviewers, Groupon businesses provide worse service than their peers, businesses discriminate by providing worse service specifically to Groupon customers, and Groupon users are less of a good fit for the businesses where they redeem Groupons.

In this paper, we return to our finding of a sharp decline in Yelp ratings scores that coincides with Groupon offers, a phenomenon we term the *Groupon effect*, and consider possible explanations through the lens of data analysis, based on an extensive dataset we gathered from Groupon and Yelp. Through this undertaking, we learn significantly more about the daily deals model, including the behavior of Groupon users and businesses. A priori, one or more of the suggested explanations might be valid, and as such we examine where data provides positive and negative evidence for each.

However, we also suggest and provide evidence for an alternative explanation that we have not heard previously. It is well known that a potential problem with review sites is that businesses may actively solicit positive reviews for their business, either through unscrupulous means such as hiring people to write positive reviews for them, or by less questionable means such as encouraging reviews from obviously enthusiastic customers. (In some cases, they may also attempt to place negative reviews for their competitors, although this is arguably a less effective strategy.) Indeed, Yelp filters its reviews to prevent "spam reviews" from affecting its ratings.

Our hypothesis is that one reason for the discrepancy in review scores is that reviews that mention Groupon correspond almost exclusively to reviews written by actual customers who use the service, and that other reviews are significantly more likely to be "fake" or otherwise introduced in an arguably artificial manner. Hence, we suggest that, at least in part, the issue is not that reviews mentioning Groupon are somehow unusually low, but that the baseline of other reviews are on average artificially high, most likely because of actions taken by businesses designed specifically to generate high-scoring reviews.

As part of our analysis efforts, we utilize a new tool for the study of the relationship between reviews and daily deal sites, namely an ordered probit analysis. Such analyses are used in regressions where the output variable is a ranked score from a finite domain, as in reviews. These regression analyses provide statistical backing for our higher-level thoughts and discussions, for which we often utilize more easily interpreted plots and graphs. Such methods have been used in other economic analyses, including in problems of adverse selection, which bear some similarity to our setting – by using Groupon as a matchmaker, businesses may be targeting customers from a less-than-ideal population, and similarly, customers are incented to select businesses from a smaller population. In both cases, customers and businesses are operating with incomplete information about each other.

After discussing related work (§ 2), and explaining our datasets and terminology (§ 3), we begin our paper by examining the evidence for the Groupon effect using extensive new data beyond that originally presented in our prior work (§ 4). We then turn to various plausible explanations in § 5, including our hypothesis regarding an excessively high baseline, and consider the evidentiary information we have to support or refute them. Along the way, various insights into Groupon, Yelp, and the interactions between these two firms arise naturally. We then deepen our analysis by interpreting our data using mathematical models and regression in § 6 to pinpoint the contribut-ing factors to the Groupon effect from key consumer-specific, merchant-specific, and review-specific attributes before concluding in § 7.

## 2. RELATED WORK

Our present work relates to multiple lines of research. First, there are a small number of recent works examining daily deal sites such as Groupon, including data analysis and models of user and business behavior. Second, past research on coupons is relevant, as a Groupon can naturally be thought of as a type of coupon. There are fairly substantial differences, however, between Groupons and traditional coupons; for example, Groupons require an up-front purchase, and they are incorporated into various social networks – one can "like" a Groupon deal on Facebook. Third, there are a significant number of papers on online review sites, covering themes such as how reviews impact purchase decisions and how reviewers' behavior can be affected by previous reviews. However, there appear to be few papers studying the type of interactions between major economic agents – in this case daily deal sites and review sites – that we examine here. As the literature on coupons and reviews related to our work is extensive, space requirements necessitate limiting ourselves to a brief discussion of relevant papers here.

The question of how daily deal sites affect the reputation of a business, and the idea of using Yelp reviews as a proxy, was initiated in [Byers et al. 2012]. Other previous works, however, had initiated studies of daily deal sites more generally, focusing primarily on the question of whether such deals would be profitable for businesses, without considering reputation [Arabshahi 2010; Dholakia 2010; Edelman et al. 2010]. Merchant reputation in the context of Yelp reviews has also been separately considered [Luca 2011].

Somewhat surprisingly, we have not found work that directly shows that coupon users provide worse reviews on online review forums or other review settings. However, there is general work on the issue of marketing efforts bringing in unprofitable customers (see, e.g., [Cao and Gruca 2005] and references therein); this can be viewed as an adverse selection problem, where marketing efforts bring in the wrong kind of customer instead of profitable new customers. It seems reasonable to extend this line of thinking to encompass the possibility that Groupons may bring in customers that provide less of a fit, and thereby worse reviews, to merchants using Groupon.

Other relevant work on coupons includes analysis that shows that customers frequently redeem coupons just before the expiration date [Inman and McAlister 1994], a useful fact that we observe in our data. Previous work showing that coupons distributed through mass media induce brand switching and reduce loyalty [Dodson et al. 1978; Shaffer and Zhang 1995] is also relevant to our study.

There has been extensive work studying online review sites that relates to our study. For example, [Chevalier and Mayzlin 2003] discusses analysis of book review sites showing that review features such as length are important to potential customers, not just the average star rating; we examine length and other features of Yelp reviews here as well. Recent work shows that online review scores tend to decline over time [Godes and Silva 2011], an issue we take into account in our analysis. Related work suggests that over time, there may be a tendency for reviews to take on extreme values [Wu and Huberman 2008; Dellarocas and Narayan 2006].

Additionally, our work makes use of a generalized ordered probit model [Terza 1985] for the regression of Yelp ratings on their characteristics. Both the generalized and the standard ordered probit model [McKelvey and Zavoina 1975] have been used extensively to model ordinal outcomes from cattle characteristics [Sy et al. 1997] to stock prices [Hausman et al. 1992].

## 3. DATASETS

Our data was collected from two websites: Groupon.com and Yelp.com. The Groupon data set consists of information about a set of 16,692 deals offered by Groupon in 20 major metropolitan areas in the US between January 3rd and July 3rd, 2011.

To collect Yelp data, we started from Groupon merchants with whom we could confidently associate a corresponding Yelp page, through one of the following two methods. First, we leveraged the fact that in many cases, the Groupon offer page contains a direct link to the Yelp page for the purpose of promoting the deal. Second, we used the search feature on Yelp with the merchant name and merchant location, and when the search resulted in an exact match, we used that association. In many instances, by searching on Yelp, we found businesses whose name and address only partially matched the corresponding business. Even though the associations were likely correct in most cases, we chose to ignore them. Through these two mechanisms, we successfully associated 5,472 Groupon businesses, a set we refer to as our *seed set*.

We then proceeded to compile two datasets of reviews from Yelp.com. For the first dataset (Dataset 1), our goal was to collect the entire review histories of *all* reviewers who had ever reviewed *any* of the businesses in the seed set prior to the collection date, January 26th 2012. Our crawl left us with 7,136,910 reviews for 942,589 distinct businesses. This dataset contains as a subset all reviews for the 5,472 Groupon businesses in our seed set.

For the second dataset (Dataset 2), we restricted attention to the review histories of the seed businesses (which constituted a subset of the reviews in our first crawl). We separated reviewers in this set into two groups: those that had never mentioned Groupon in any of their reviews (127,946 reviewers), and those that had mentioned Groupon at least once (21,020 reviewers). The latter set of reviewers had reviewed 464,706 businesses other than those in the seed set. We collected the complete review histories of a random sample of 25,128 of these businesses (many of which have presumably never run a Groupon offer).

We note that Yelp algorithmically *filters* reviews based on legitimacy and how "established" the reviews' authors are (see [Yelp Web Blog 2010] for details). Filtered reviews do not count towards Yelp ratings, and filtered reviews for a merchant are not visible to users by default, although they can be accessed by solving a CAPTCHA. A review's status is not permanent: as users become more or less established, their reviews can switch from being visible to being filtered and vice-versa. Yelp purposefully avoids elaborating precisely what factors determine how established users are and whether reviews are filtered to limit fraud.

To be clear, review histories of *users* contain *both* their filtered and their unfiltered reviews, and no indication is given as to which reviews are filtered. (To make this determination, one could cross-check a review to see if it is visible on the *merchant* page.) Hence Dataset 1 contains filtered reviews, while our methodology for Dataset 2 collected only those reviews that were not filtered at the time of collection.

### 3.1. Terminology

We employ the following terminology to describe certain aspects of our dataset in the paper. We use the term *Groupon business* to refer to any merchant that ran a Groupon deal in our seed set. We will use the term *Groupon review* to refer to any Yelp review for a Groupon business that contains the word "Groupon" in its text and we will call the author of one or more such reviews a *Groupon user*. We will use the terms *non-Groupon business*, *non-Groupon review*, and *non-Groupon user* to refer to the corresponding complementary sets.

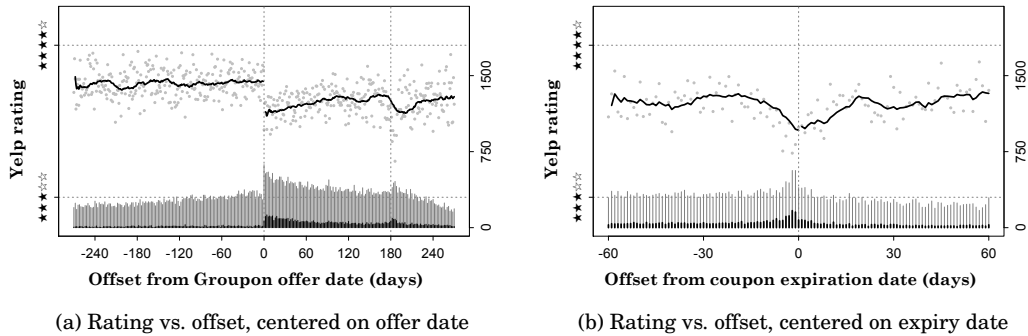

(a) Rating vs. offset, centered on offer date          (b) Rating vs. offset, centered on expiry date

Fig. 1: Yelp review scores and volumes for Groupon merchants, centered on Groupon offer date and Groupon expiration date, respectively

A review on Yelp.com consists of a star rating, and some free text. The star rating takes on a value from $\{1, 2, 3, 4, 5\}$. For each Groupon business, we associate an offer date that corresponds to the date they initiate a Groupon offer. Then, for every review of a Groupon business, we associate an integer offset with that review reflecting how many days after (or before) the offer date the review was posted. For example, a review posted on March 7th for a business that subsequently initiates a Groupon offer on March 13th would have an offset of -6.

## 4. REVIEW OF THE GROUPON EFFECT

We begin by reviewing evidence and providing new evidence for the finding that Groupon offers coincide with substantially lower ratings for Groupon businesses than other reviews, and that this is caused by Groupon users. The most telling evidence comes from comparing mean ratings from Groupon reviews and non-Groupon reviews for our seed set: Groupon reviews have a mean score of 3.27 stars, while non-Groupon reviews have a mean of 3.73 stars. This discrepancy is somewhat larger than what we initially reported in [Byers et al. 2012] on a smaller data set. We can gain more insight into the effects of Groupon offers via some simple visualizations.

**Discontinuities at the Groupon offer date:**  In Figure 1a, the top scatterplot and trend line capture the relationship between the average Yelp rating and the offset for reviews of Groupon businesses. Each point records the average rating of all reviews with a given offset across all Groupon businesses, using the Yelp star rating as depicted on the left side of the y-axis. The discontinuities seen at offset zero coincide with the Groupon offer date. The trend lines are computed as a 30-day moving average across offsets, with the average resetting at offset zero to highlight the different behavior at the Groupon date. (From offset 0, only $k + 1$ days are averaged at offset $k$, and similarly at the left end of the plot.) The histograms at bottom reflect the daily review volume for each given offset, using the scale on the right side of the y-axis for the number of reviews. The smaller histograms with darker shading reflect the volume of Groupon reviews (i.e. those mentioning Groupon specifically). Again, there are striking discontinuities at offset zero as review volumes surge subsequent to the Groupon offer. Note that Groupon reviews account for only about half of the increase, suggesting there exist Groupon users who do not mention Groupon in their review. Finally, observe the gradual increase in review volume prior to offset zero: this is consistent with the rapid rise in popularity of Yelp as a review site, and we view it as independent of effects due to Groupon.

As mentioned, in prior work, we noted a decline in Yelp rating scores coincident with a Groupon offer [Byers et al. 2012]. However, in that plot (reproduced for reviewers as Figure 6 in the Appendix), the evidence is much more ambiguous, in large part because the data was presented at a monthly granularity. At that granularity, and on a much smaller dataset than that studied in this paper, skeptical observers suggested that review scores appeared to be declining even *before* the Groupon offer, and seem to continue to trend downward thereafter. Our more detailed plot in Figure 1a clarifies that the decline in Yelp rating scores coincides precisely with the Groupon offer date.

**Discontinuities at the Groupon expiration date:** Careful inspection of Figure 1a reveals some interesting additional trends. Notably, there is a slow recovery of the average rating score during the Groupon offer period, in approximately direct inverse proportion to the number of Groupon reviews, which is consistent with a model in which Groupon users are the primary culprit dragging down review scores. But around offset 180, we see a pronounced drop in rating scores accompanied by a smaller secondary surge in reviews. We attribute this event to Groupon users with expiring coupons, noting that 52% of the deals in our dataset are valid for approximately 6 months (between 180 and 190 days). In Figure 1b, we again plot review scores, but renormalize the offset of each review with respect to the offer expiration date recorded in our dataset. We observe that as the Groupon expiration date nears, there is an increase in both reviews and Groupon mentions, consistent with previous work showing increased coupon redemption near the expiration date [Inman and McAlister 1994]. Moreover, as the expiration date nears, there is another sharp decline in review scores, that corrects itself subsequent to expiry. We believe this provides further evidence of the negative impact that Groupon users have on Yelp review scores.

## 5. THE GROUPON EFFECT: A ROOT CAUSE ANALYSIS

We now investigate *why* these marked discontinuities occur. We have compiled many hypotheses and classified them as follows:

— **Intrinsic Decline:** It is well known that review scores fall over time, and this is the effect seen (largely independent of Groupon).
— **Critical Reviewers:** Groupon users are more critical than their peers, and thus marketing to them is a case study in *adverse selection*.
— **Bad Businesses:** Merchants who feel compelled to offer a Groupon are desperate, or in trouble anyway. It's no wonder they get poor reviews.
— **Unprepared Businesses:** The surge of customers overwhelms many merchants, and the quality of service they offer falls as a result.
— **Discriminatory Businesses:** Groupon businesses actively discriminate against customers with a Groupon, and selectively deliver lower-quality service to those individuals. If Groupon users are unaware of this discrimination prior to purchase, this is another instance of adverse selection: consumers adversely selecting businesses.
— **Experimentation:** Groupon users are often experimenting when they purchase a Groupon. As a result, they tend to have a more tenuous fit with the merchant, resulting in lower scores on average.
— **Artificial Reviews:** Finally, we offer our own novel hypothesis: Groupon reviews are actually a more accurate and realistic baseline, because the rest of the reviews contain a higher fraction of artificially laudatory reviews that may have been incentivized or entered fraudulently. Non-Groupon review scores are inflated as a result.

We next consider the evidence from our data for these hypotheses in turn.

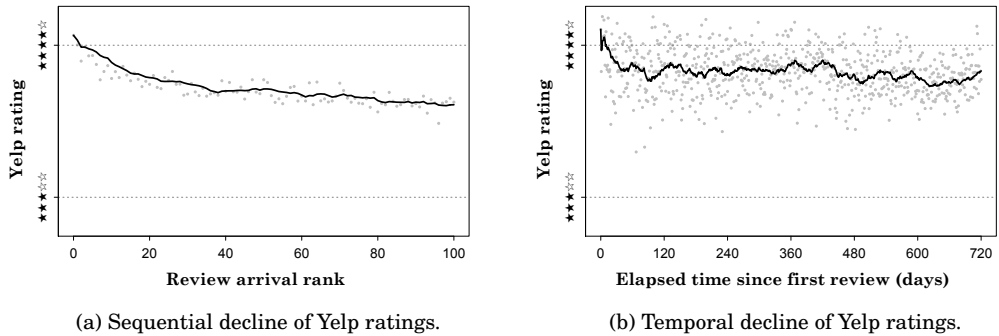

(a) Sequential decline of Yelp ratings.          (b) Temporal decline of Yelp ratings.

Fig. 2: The intrinsic decline of Yelp ratings by arrival rank and over time.

## 5.1. Hypothesis: Ratings suffer an intrinsic decline over time

One common explanation we heard for declining review scores is completely independent of Groupon: that review scores naturally decline over time. Empirical studies examining review data as a time-series support this claim, and models have been proposed for explaining this phenomenon. For example, work by Godes and Silva empirically analyzes these effects for review data from Amazon.com on top-selling books [Godes and Silva 2011]. They consider two closely related, but distinct, attributes for each book review: the elapsed time (measured in days) since the *first* review for that book appeared, and the sequential arrival rank of each review (i.e. arrival rank 5 corresponds to the fifth review written for that book). Their data supports modeling trends with respect to each of these attributes differently. In contrast, we observe similar phenomena for both attributes. In Figure 2a we plot the average rating of reviews grouped by their arrival rank (limiting our study to 663 businesses in our dataset with at least 100 reviews each). Similarly, Figure 2b displays the average daily rating for the 3,534 business in our dataset first reviewed on Yelp at least 2 years prior to our data collection. Qualitatively, we observe a decline on average during the earliest phase of Yelp reviewing (whether by time or arrival). However, the significant declines during initial review period are not sustained, and the much more gentle subsequent decline cannot explain our Groupon-based observations. We formally quantify the influence of temporal factors from a regression modeling standpoint in § 6.

## 5.2. Hypothesis: Groupon users are more critical

From the review data, it is clear that the opinions of Groupon users about Groupon businesses differ substantially from those of their peers. A natural hypothesis is that Groupon users are *in general* more critical than their peers, as demonstrated by lower rating scores, in which case the decline in scores after a Groupon offer might be attributable to this factor.

We investigate the question of whether one class of reviewers is generally more critical than another. A more critical class of reviewers would review *all* businesses more critically on average, irrespective of whether they ran a Groupon. The fairest comparison to isolate this effect is to compare evaluation on non-Groupon businesses. We classify reviewers into Groupon reviewers and non-Groupon reviewers (as described in Section 3.1). In Figure 3a, we plot histograms of the star rating scores of all 274,416 reviews by Groupon reviewers (darker shading) and 1,605,571 reviews by non-Groupon reviewers (lighter shading) for non-Groupon businesses. (These reviews are determined by business, and as such do not contained filtered reviews.) The results are surprising. We do not see strong evidence that Groupon reviewers are much more critical on average. Instead, they are more *moderate*! They provide fewer 1 and 5-star ratings than their non-Groupon counterparts and more 3 and 4-star ratings.

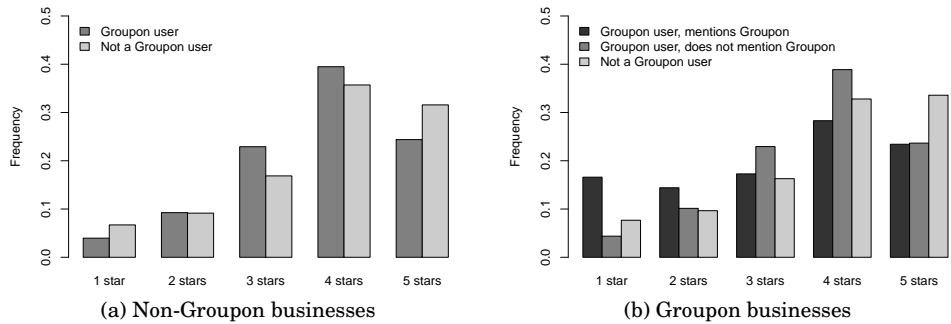

Fig. 3: Empirical distribution of Yelp star-ratings by user and by review type.

We also investigate differences in reviews for Groupon businesses. We classify reviews for Groupon businesses into three categories, and plot the resulting histograms in Figure 3b. The categories are: reviews by non-Groupon users (lightest shading), reviews by Groupon reviewers that contain a Groupon mention (darkest shading); and reviews by Groupon users who do not mention Groupon for that business, but mentioned Groupon elsewhere (medium shading). The histograms for the reviews that do not mention Groupon are quite similar to the corresponding histograms for non-Groupon businesses in Figure 3a in terms of how Groupon users and non-Groupon users behave. The histograms for Groupon users (medium and dark shading), interestingly, exhibit prominent differences: while the frequency of 5-star reviews are comparable in both classes, significant mass moves from 4- and 3-star reviews to 2-star and 1-star reviews. Our interpretation is that this evidence strongly points not to critical reviewers, but to unrealized expectations between reviewers and merchants. Moreover, while this mismatch appears to occur relatively infrequently, when it does occur, the result is a much more negative review than is typical for that reviewer.

Later in the paper, we will return to other reviewer attributes besides criticality that more clearly distinguish Groupon reviewers and non-Groupon reviewers. To get a flavor for the distinctions, consider the summary statistics of Groupon user profiles on Yelp vs. non-Groupon user profiles in Table I. While both classes have roughly the same tenure on Yelp, the Groupon users have more friends, more fans, and write more reviews. We later show that Groupon users' reviews are also longer, and more critically acclaimed. The bottom line is that while Groupon reviewers are not significantly more critical, they are *mavens* in the Gladwell sense [Gladwell 2000], and as such their reviews appear to carry more weight than those of their non-Groupon peers. It is worth mentioning that [Ghose and Ipeirotis 2010] finds that more informative reviews, even if they provide negative ratings, can *increase* products sales.

### 5.3. Hypothesis: Groupon Businesses are More Likely to be "Bad" Businesses

In our classification of hypotheses, there were several pointing the blame at merchants: merchants might be unprepared, in trouble, or discriminatory against Groupon-bearing customers. Across such a large dataset, we would expect some number of merchants of each type, and indeed, the popular press excels at highlighting negative anecdotal evidence, amplifying the sense that "bad" merchants may be common. One prominent incident was a promotion that Groupon ran with FTD Flowers in February 2011. Customers who purchased the Groupon soon realized that when they visited the site, they were shown (and charged) much higher prices for the same products than customers without Groupons. This deal was widely viewed as a "bait and switch" scheme [Techcrunch 2011], and Groupon offered full refunds, but not without suffering a reputational hit.

Table I: Summary statistics of user profiles for Groupon users vs. non Groupon users. Means and standard deviations (in parentheses).

|                   | Yelping Since | Friends  | Fans    | Reviews  | Firsts  | Count   |
|-------------------|---------------|----------|---------|----------|---------|---------|
| Groupon user      | 2009-06-27    | 44.94    | 4.38    | 89.60    | 7.19    | 21,020  |
|                   | (506.18)      | (144.28) | (16.74) | (160.34) | (29.40) |         |
| Not a Groupon user| 2009-06-01    | 24.43    | 1.92    | 44.25    | 3.72    | 127,946 |
|                   | (530.01)      | (106.62) | (12.49) | (88.57)  | (19.32) |         |

To test hypotheses about bad businesses rigorously, we conducted various experiments. Some, such as a text-based sentiment analysis of Yelp reviews that we performed to try to cluster the underlying reasons for negative Groupon reviews, did not bear fruit. But conducting statistical tests for a preponderance of "bad businesses" amongst the Groupon businesses was more enlightening. Our approach was as follows: first, we know already that on average, reviews for Groupon business fall during the offer period. But if we observe a skewed distribution of rating score declines across Groupon businesses (relative to a baseline), this supports the hypothesis of a subset of bad Groupon businesses accounting for a disproportionate share of the decline.

To compare observed declines against a baseline, we started with a control group: we considered the review scores of a set of 4,037 non-Groupon businesses in a 90-day before and a 90-day after period around an arbitrary date (2011-04-01), limited to businesses that had received at least 10 reviews in each of the periods. For each merchant $j$, we computed the difference between average Yelp rating scores for that merchant in the before period vs. the after period, and plotted the frequencies of these scores (light shading) in Figure 4a. Not surprisingly, these differences are quite close to a normal distribution, but with higher kurtosis: they have mean -0.02, variance 0.15, and slight negative skew -0.01, consistent with review declines over time observed earlier.

Next we considered the experimental group, Groupon businesses. Again we computed the before and after difference in rating score, but this time the periods were centered on the Groupon offer date. The corresponding frequencies for these businesses appear in dark shading in Figure 4a. We observe the expected translation effect to mean $-0.12$ (and variance $0.18$) reflecting the negative impact of a Groupon offer, but more interestingly, the sample skewness $-0.05$ is significantly greater than that of the control group. This supports the hypothesis that some of the Groupon businesses are disproportionately responsible for negative reviews than others.

We also examined the rating scores of Groupon businesses on a per-business basis, comparing Groupon reviews to non-Groupon reviews. For each business $j$, we compute the difference between the average Groupon-mentioning rating and the average non-Groupon rating for that business. The frequencies of these scores are plotted in Figure 4b. Interestingly, this distribution also has high negative skew, perhaps indicative of discrimination against Groupon customers by a small subset of merchants.

We further investigated whether certain business categories or deal attributes have stronger associations with negatively skewed reviews. In general, our analyses sug-

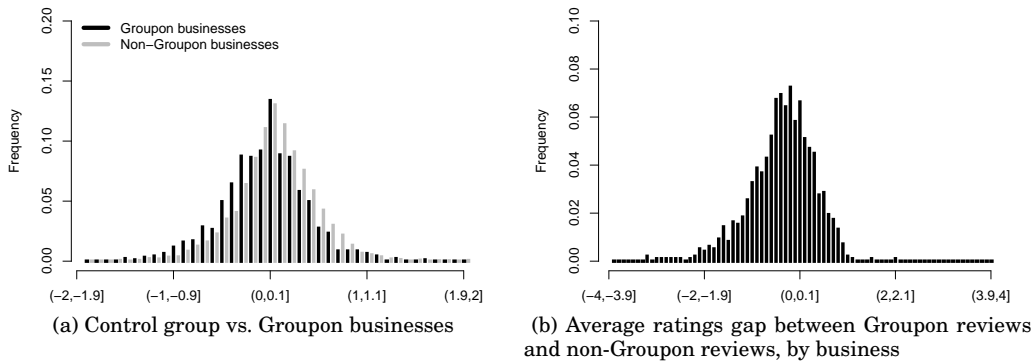

(a) Control group vs. Groupon businesses

(b) Average ratings gap between Groupon reviews and non-Groupon reviews, by business

Fig. 4: Empirical distribution of ratings deltas.

gested differences across categories, but no one especially noteworthy finding. A representative experimental finding is depicted in Table X (Appendix), where we consider the magnitude of decline across categories with at least 100 deals. While the magnitudes of the declines vary, the differences are not especially significant.

To conclude, we believe that the "bad business hypothesis" is backed by empirical evidence, but only to a limited extent. While we see some change in the shape of the distribution of the change in Yelp scores between Groupon and non-Groupon businesses, we do not see the extreme differences that would explain the large difference in review scores. We provide more robust evidence for this conclusion by regressing against both business-specific factors and customer-specific factors in § 6.

### 5.4. Hypothesis: Groupon users are often engaging in experimentation

Our next class of hypotheses considers users and merchants in tandem. Our findings in Section 5.2 demonstrated that on occasion, but with atypical frequency, a Groupon user visits a merchant and writes a highly negative review. This mismatch of expectations could arise from a "bad" business, as described in the preceding section, or it could arise from a mismatch in expectations between user and merchant. One hypothesis (well-known in the coupon literature, see e.g. [Bawa and Shoemaker 1987; Raju 1992; Shaffer and Zhang 1995]) is that coupons encourage experimentation, and indeed, the significant Groupon discounts intuitively support this hypothesis.

To test for experimentation, we applied the following methodology to quantify the degree of experimentation for reviews within a Yelp reviewer's complete review history. First, we note that each Yelp review is associated with up to three specific business categories (out of a list of hundreds), such as "Peruvian", or "Roofing". Our first test classifies a review as experimental if the reviewer's history contains zero other reviews for businesses that share any common category with the reviewed business. Our second test uses the zip code for each Yelp business as the discriminant. A review was deemed to be experimental if no other review in that review history was from a business in the same zip code. We restricted attention to reviewers who reviewed at least one business in the seed set and who also had significant review histories, comprising at least 10 reviews.

We now follow the segmentation of reviews into the three categories that we used when studying critical reviewers to investigate which classes of reviewers conduct the most experimentation. Our findings are presented in Table II. Rows 2 and 3 are the most striking: they indicate that Groupon users are much more likely experimenting when they are using a Groupon than when they are not. Row 1 indicates that non-Groupon users also significantly engage in experimentation, apparently more than Groupon users. However, there is a significant caveat: non-Groupon users write fewer reviews on average than Groupon users (recall Table I), and our simple measure of experimentation presented here does not normalize for size of review history[1]. We include review history later in our regression analysis to account for its effect, but we note here the effect appears fairly small.

Table II: Summary statistics of customer experimentation.

| Groupon user | Groupon mention | Category match? | | ZIP match? | |
| --- | --- | --- | --- | --- | --- |
| | | Yes | No | Yes | No |
| False | False | 70% | 30% | 68% | 32% |
| True | False | 84% | 16% | 80% | 20% |
| True | True | 67% | 33% | 66% | 34% |

Ultimately, the results we obtain support the hypothesis that Groupon users engaging in experimentation more often when they use a Groupon. In an extension to our basic regression model that we describe in § 6.4, we conduct bivariate regressions to formally capture the extent to which experimentation correlates with lower Yelp review scores.

### 5.5. An Artificial Baseline

As noted in the introduction, we suggest a new hypothesis explaining lower Groupon scores: other scores are actually artificially high. One approach to test the artificial baseline hypothesis is to consider filtered reviews. Recall that Yelp algorithmically filters reviews that it considers less trustworthy using proprietary methods. Hence we start by determining if Groupon reviews on the whole are more or less trustworthy according to Yelp.

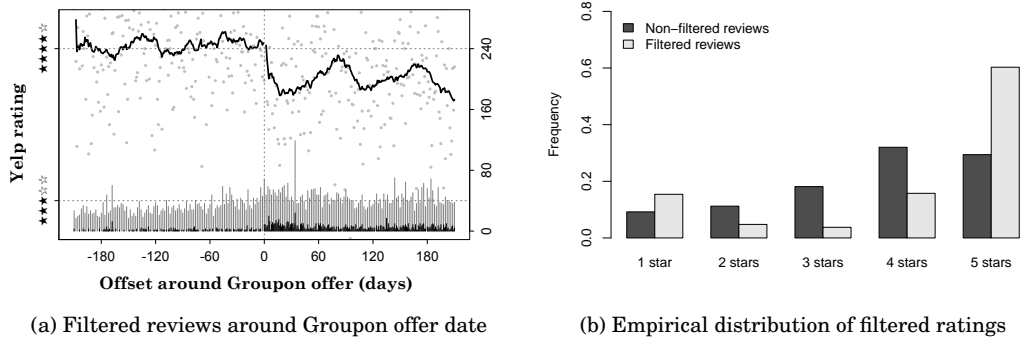

(a) Filtered reviews around Groupon offer date    (b) Empirical distribution of filtered ratings

Fig. 5: Rating scores and volumes of filtered Yelp reviews for Groupon merchants, viewed as a time-series and as a histogram.

---

[1] All other things being equal, larger review histories naturally tend to span more Yelp categories and ZIP codes, and thus any individual review within that history is less likely to be classified as experimental.

Table III: Percentage of filtered reviews for Groupon vs. non-Groupon users.

| Groupon user | Groupon mention | Reviews | | | |
|---|---|---|---|---|---|
| | | Visible | Filtered | Filtered pct. | Avg. Rating |
| False | False | 4,837 | 723 | 14.95% | 3.79 |
| True | False | 6,496 | 707 | 10.88% | 3.58 |
| True | True | 175 | 19 | 10.86% | 3.28 |

We performed the following experiment: starting from the set of all reviewers in our seed set, we selected 100 random Groupon users, and 100 random non-Groupon users. According to the users' review histories, they had jointly reviewed 9,477 businesses. To check whether the users' reviews had been filtered, we collected all reviews for the corresponding business pages (recall that filtered reviews appear in user histories, but *not* on the corresponding business page). This amounted to 1,835,734 reviews over all 9,477 businesses. (Clearly this test for filtering is not especially scalable; we hope to devise a more scalable test in future work.) The results shown in Table III indicate that non-Groupon users' reviews are substantially more likely to be filtered; moreover, filtered review scores from Groupon users are unsurprisingly significantly lower than for non-Groupon users. Given that reviews from non-Groupon users are more common, we believe that these results along with the distributions shown in Figure 3 provide evidence that artificially high reviews have a non-trivial effect when comparing Groupon and non-Groupon reviews.

As a related issue, Groupon users in fact appear to give higher quality reviews than the average reviewer, suggesting further that Groupon reviews are less likely to be artificially inflated. We use the length of a review and other users' ratings of reviews as a proxy for realism and usefulness. The length of a review provides some measure of its quality. More directly, Yelp lets subscribers mark other reviews as Funny, Useful, or Cool. Summary statistics for our three classes of reviews are presented in Table IV. Reviews by Groupon users are in fact longer than average, and much more likely to be flagged as useful by their peers. Strikingly, their reviews are perceived as much funnier and cooler when they are writing reviews that do not mention Groupon, but even the Groupon reviews are still relatively funny and cool. While the connection is arguably circumstantial, we believe these metrics demonstrate that Groupon users provide a baseline of high-quality reviews and scores; assuming this is an accurate assessment, it provides further evidence that non-Groupon review scores are inflated by low-quality and apparently biased reviews.

Our final set of observations on filtered reviews are compiled in Figure 5. In Figure 5a, we plot the average Yelp rating and review volume of filtered reviews, centered around the Groupon offer date, with the trend line reflecting a 30-day moving average. We observe two noticeable differences in comparison to Figure 1a. First, the surge of Groupon-mentioning reviews coincident with the offer date is markedly less dramatic. Second, the rating score decline is less pronounced. Our interpretation is that fewer Groupon-related reviews are being filtered. In Figure 5b, we compare the empirical distributions of ratings for filtered and unfiltered reviews. As expected, we see that most filtered reviews have highly polarized review scores: nearly all provide 1-star, 4-star, or 5-star reviews.

## 6. MODELING THE GENERATION OF YELP RATING SCORES

We now describe mathematical models that we employ to frame our statistical analysis of our datasets. Our main approach is a generalized ordered probit model [McKelvey and Zavoina 1975; Terza 1985].

### 6.1. Ordered Probit

Consider the following model for the latent process driving the generation of a Yelp rating:

$$y_{ij}^* = \mathbf{x}_{ij}'\boldsymbol{\beta} + \epsilon_{ij}, \tag{1}$$

where $y_{ij}^*$ is a continuous dependent random variable on $(-\infty, \infty)$ representing consumer $i$'s unobserved opinion of a business $j$, $\mathbf{x}_{ij}'$ is a vector of covariates, $\boldsymbol{\beta}$ is a vector of coefficients, and $\epsilon_{ij}$ is an random disturbance. We assume that the observed Yelp star ratings $y_{ij} \in \{1, 2, 3, 4, 5\}$ are determined by the $y_{ij}^*$ via thresholds. Hence

$$y_{ij} = \begin{cases} 1 & \text{if } y_{ij}^* < \kappa_1, \\ 2 & \text{if } \kappa_1 \leq y_{ij}^* < \kappa_2, \\ 3 & \text{if } \kappa_2 \leq y_{ij}^* < \kappa_3, \\ 4 & \text{if } \kappa_3 \leq y_{ij}^* < \kappa_4, \\ 5 & \text{if } \kappa_4 \leq y_{ij}^*, \end{cases} \tag{2}$$

where $\kappa_1 < \kappa_2 < \kappa_3 < \kappa_4$ are constant thresholds. Notice that the model places no further restriction on the thresholds, such as assuming the difference between a 1-star rating and a 2-star rating is the same as the difference between a 4-star and a 5-star rating. Under the assumption or normally distributed random noise with known variance, $\epsilon_{ij} \sim \mathcal{N}(0, 1)$ we have

$$Pr[y_{ij} \leq n] = \Phi(\kappa_n - \mathbf{x}'\boldsymbol{\beta}) \tag{3}$$

and therefore

$$Pr[y_{ij} = n] = \Phi(\kappa_n - \mathbf{x}_{ij}'\boldsymbol{\beta}) - \Phi(\kappa_{n-1} - \mathbf{x}_{ij}'\boldsymbol{\beta}) \tag{4}$$

where $\Phi$ is the standard normal distribution function. (Fixing the variance of the errors is an identification constraint of the ordered probit model. The variance cannot be estimated jointly with all the other parameters of the model.) These basic model assumptions correspond to the classical ordered probit model [McKelvey and Zavoina 1975], which can be estimated straightforwardly using maximum likelihood.

Table IV: Summary statistics of reviews of Groupon businesses. Means and standard deviations (in parentheses).

| Groupon user | Groupon mention | | Votes | | | | |
| | | Length | Funny | Useful | Cool | Stars | Count |
|---|---|---|---|---|---|---|---|
| False | False | 715.15 | 0.23 | 0.51 | 0.27 | 3.76 | 176,653 |
| (*Reviewer has never mentioned Groupon*) | | (608.96) | (1.30) | (1.74) | (1.21) | (1.24) | |
| True | False | 881.01 | 0.62 | 1.19 | 0.73 | 3.68 | 37,453 |
| (*Groupon user, mentioned Groupon elsewhere*) | | (662.49) | (2.21) | (2.66) | (2.12) | (1.09) | |
| True | True | 1109.48 | 0.37 | 1.11 | 0.37 | 3.28 | 13,602 |
| (*Groupon user, mentioned Groupon*) | | (771.17) | (1.43) | (2.80) | (1.34) | (1.40) | |

### 6.2. Generalized Ordered Probit and the Parallel Regressions Assumption

The constant thresholds of the classical ordered probit model above implicitly make the so-called parallel regressions assumption. The regressions are "parallel" in the sense that the coefficients $\beta$ do not vary across ordinal rating values. This assumption is restrictive and often violated in practice. Terza suggests a generalization of the model which relaxes this assumption [Terza 1985], allowing the thresholds to be linear functions of some of the covariates:

$$Pr[y_{ij} \leq n] = \Phi(\kappa_n - \mathbf{w}'_{ij}\boldsymbol{\gamma}_n - \mathbf{x}'_{ij}\boldsymbol{\beta}), \text{ for } n = \{1, \ldots, 4\}, \tag{5}$$

where $\mathbf{w}_{ij}$ contains the covariates with coefficients unconstrained across different values of $n$, and $\mathbf{x}_{ij}$ contains those with parallel coefficients. Every explanatory variable in the model is parameterized by either a single $\beta$ coefficient (if it is constrained in parallel), or four $\gamma$ coefficients if it is unconstrained across rating values. Notice that the standard ordered model is a special case of the generalized model, and further suggests a likelihood ratio for testing the parallel regressions assumption. Our findings, given in the Appendix, strongly suggest that the parallel regressions assumption is violated, as expected from our observations in prior sections, and thus we use the generalized model henceforth.

### 6.3. Factors underlying the latent review function

We now focus on the aspects of consumer behavior and their experience with the merchant that drives their evaluation of a business, resulting in their latent review score. Our model for the latent review function leverages our prior observations to attribute scores to three additive factors: 1) consumer-specific effects, due to their direct experience with the merchant; 2) business-specific effects, including their location and whether they are running a Groupon offer; and 3) Yelp-specific effects, principally effects due to review arrival order. The use of a latent response function to model consumer satisfaction in the context of online reviews, as well as the distinction of factors underlying it has been considered in prior work [Ansari et al. 2000; Ying et al. 2006; Moe and Schweidel 2011]. Our work departs from these prior works in using a generalized ordered probit model to relax the parallel regressions assumption and by computing per-rating value coefficients. In contrast, prior work often deals with the issue by employing random consumer and business effects, essentially allowing the coefficients to vary by individual and business under a distributional assumption (usually taken to be a multivariate normal distribution on the joint distribution of the coefficients).

We model the latent response function as follows:

$$\text{probit}(\text{Pr}[y^*_{ij} \leq n]) = \kappa_n - C_{in} - B_{jn} - R_{ijn}, \tag{6}$$

with consumer $C_{in}$, business $B_{in}$, and review effects $R_{ijn}$ taken to be

$$C_{in} = \gamma_{1n} \times \text{Groupon user}_i,$$

$$B_{jn} = \sum_{p=2}^{\#cities} \beta_{2p} \times \text{Deal city}_j + \sum_{q=2}^{\#categ.} \beta_{3q} \times \text{Deal category}_j$$

$$+ \gamma_{2n} \times \text{During Groupon}_j + \gamma_{3n} \times \text{Post Groupon}_j,$$

$$R_{ijn} = \gamma_{3n} \times \text{Groupon mention}_{ij} + \gamma_{4n} \times \text{Review rank}_{ij}.$$

Variables prefixed with $\beta$ coefficients are constrained in parallel, and those prefixed with $\gamma$ coefficients are not. We chose to constrain deal city and category for parsimony and to avoid calculating spurious effects. Fitting the fully unconstrained model did not result in any significant change in our results. With the exception of the Review rank$_{ij}$, which take on positive integer values, the variables in the model are either binary or dummy-coded. We chose to model the temporal decline of reviews using arrival rank instead of elapsed time for two reasons: 1) the two effects are highly correlated in our dataset, and 2) due to large volumes of Groupon reviews, time can artificially be compressed, and thus we anticipate that arrival rank may better model rating declines. The variables During Groupon$_j$ and Post Groupon$_j$ are mutually exclusive and are used to capture temporal effects with respect to the period prior the starting date of deal $j$. Deal city and category are computed relative to notional reference levels (Atlanta, and Arts & Entertainment respectively). Groupon user$_i$ and Groupon mention$_i$ are indicators with the obvious meaning, but if the latter indicator is set to 1, so is the former, and as such their coefficients have to be interpreted jointly in that case.

Table V: Coefficient estimates for the generalized ordered probit regression of Yelp ratings. Standard errors are shown in parentheses. Log-likelihood -316,365, AIC 632,846. Business city and category coefficients are omitted for brevity.

| | probit($Pr[y_{ij} \leq n]$) | | | |
|---|---|---|---|---|
| | n=1 | n=2 | n=3 | n=4 |
| (Intercept $\kappa_n$) | -1.48 (0.019) | -1.01 (0.018) | -0.49 (0.018) | 0.37 (0.018) |
| Groupon mention | -0.68 (0.017) | -0.54 (0.014) | -0.32 (0.013) | -0.13 (0.014) |
| Groupon user | 0.29 (0.012) | 0.13 (0.008) | -0.10 (0.008) | -0.28 (0.008) |
| During Groupon deal | -0.24 (0.009) | -0.17 (0.007) | -0.05 (0.007) | 0.06 (0.007) |
| Post Groupon deal | -0.19 (0.012) | -0.13 (0.010) | -0.05 (0.009) | 0.04 (0.009) |
| Review rank | 3.8E-4 (3E-5) | 1.6E-4 (2E-5) | 0.9E-4 (2E-5) | 3.6E-4 (2E-5) |

We estimated the coefficients for Yelp star rating against the observed independent variables for our model using the VGAM library for R [Yee 2010]. The results are shown in Table V. To interpret the model's coefficients, note that for a given variable (row) and a given value of $n$ (column), a positive coefficient in the corresponding cell implies that an increase in the value of that variable yields a *decrease* in the corresponding probability $Pr[y_{ij} \leq n]$, in accordance with Equation 5. For example, consider the entry for "During Groupon deal" for $n = 2$. The negative value $-0.16696$ here reflects the increased likelihood that a review during this timeframe will be a 1- or 2-star review.

Indeed, an alternative and easier way to interpret the model involves computing marginal effects. Recall the marginal effect with respect to the $m^{th}$ explanatory variable is defined as $\partial P[y_{ij} = n|x_{ij}]/\partial x_{ij}^{(m)}$. Similarly, the expected marginal effect is then defined to be

$$E_x \left[ \frac{\partial P[y_{ij} = n|x_{ij}]}{\partial x_{ij}^{(m)}} \right],$$

a quantity which can be consistently estimated by averaging over all observations in our dataset. The values of the marginals are provided in Table VI.

For example, looking the average marginal effects for "Groupon user" we see that, all else equal, the ratings of Groupon users are more moderate: they are less likely to provide both 1- and 5-star reviews and more likely to provide 2-, 3-, and 4-star reviews, something that we have already empirically observed in Figure 3b. As mentioned, the coefficients for "Groupon mention" have to be interpreted jointly with the coefficients for "Groupon user" by definition (we can think of "Groupon mention" as an interaction effect on "Groupon user"). Hence a Groupon user who mentions Groupon is much more likely to give a 1-star rating and substantially less likely to give a 5-star rating.

Table VI: Average marginal effects of the explanatory model variables on probalities of receiving a specific Yelp rating.

|  | Yelp rating | | | | |
|  | 1 | 2 | 3 | 4 | 5 |
| --- | --- | --- | --- | --- | --- |
| Groupon mention | 10.223% | 3.802% | -2.242% | -7.388% | -4.394% |
| Groupon user | -4.436% | 1.095% | 6.544% | 6.398% | -9.601% |
| During Groupon deal | 3.577% | 0.778% | -2.544% | -4.017% | 2.206% |
| Post Groupon deal | 2.791% | 0.575% | -1.737% | -3.069% | 1.440% |
| Review rank | -0.006% | 0.002% | 0.008% | 0.010% | -0.013% |

Table VII: Estimates for the bivariate probit regression (Equation 8) on Yelp matching category and ZIP. Log-likelihood -223024.8, AIC 446199.7, $\rho$ = 0.31 (Std. Err. 0.003). Business city and category coefficients are omitted for brevity.

|  | Matching category | | | Matching ZIP | | |
|  | Est. | Std. Err. | $t$-value | Est. | Std. Err. | $t$-value |
| --- | --- | --- | --- | --- | --- | --- |
| (Intercept) | -0.46 | 0.02 | -18.94 | -0.18 | 0.02 | -7.83 |
| Groupon user | 0.11 | 0.01 | 11.48 | 0.05 | 0.01 | 6.09 |
| Groupon mention | -0.10 | 0.02 | -6.13 | -0.10 | 0.01 | -6.57 |
| log(Review count) | 0.58 | 0.002 | 231.64 | 0.45 | 0.002 | 204.53 |

Looking at the table values for "During Groupon deal" and "Post Groupon deal", which are relative to the default, pre-Groupon deal, we see that there are some lasting effects after the Groupon deal, but the values appear to be regressing toward the pre-Groupon state (which would be reflected by all 0 coefficients). Our available data on the post-Groupon period is limited, but we might expect this trend to continue farther out in time, if we believe the effects of the Groupon deal on reviews fade and the business returns to its pre-Groupon status quo. One surprising finding involves the small increase in 5-star ratings during the Groupon deal. We offer two potential explanations: this could be due to a subset of truly enthusiastic reviewers, happy with their Groupon experience but failing to mention Groupon in their reviews. An alternative but intriguing explanation is offered by prior work that has argued that the incentive to post is related to the polarization of the review environment [Engel et al. 1973]. Here, customers with 5-star experiences have greater incentive to post in a negative review climate, as it amplifies the impact of their opinion. More cynically, the incentive for a merchant to solicit artificial laudatory reviews in a negative review climate is also greater. Finally, the review rank appears to have a negligible effect in our model, demonstrating that simple passage of time does not go far to explain the Groupon effect, as we have previously argued.

### 6.4. A bivariate probit model of consumer experimentation

We now consider the specific impact of Groupon usage on consumer experimentation, with a separate regression model.

Table VIII: Predicted probabilities for the bivariate probit regression (Equation 8).

| Groupon user | Groupon mention | $(Y_{cat}, Y_{zip}) = (1, 1)$ | $(1, 0)$ | $(0, 1)$ | $(0, 0)$ |
|---|---|---|---|---|---|
| False | False | 38.49% | 15.65% | 14.54% | 31.30% |
| True | False | 66.92% | 14.05% | 9.09% | 9.23% |
| True | True | 34.73% | 15.77% | 14.13% | 35.35% |

Given a Yelp review, let $Y_{cat}$, $Y_{zip}$ be random binomial variables indicating the existence of another review with a matching Yelp category or ZIP code, respectively, within a user's review history. Following our probit methodology let

$$\Pr[Y_{cat} = 1|\mathbf{x}] = \beta_1\mathbf{x} + \epsilon_1 \ ; \ \ \Pr[Y_{zip} = 1|\mathbf{x}] = \beta_2\mathbf{x} + \epsilon_2, \qquad (7)$$

where $\mathbf{x}$ is a vector of covariates, $\beta_{\{1,2\}}$ are vectors of coefficients, and $\epsilon_{\{1,2\}}$ are random disturbances. Assuming again that $\epsilon_{\{1,2\}} \sim \mathcal{N}(0,1)$, this leads to two independent probit models that we can estimate. Of course, common sense indicates that the decisions to geographically explore and to try something new are correlated; therefore, separate estimation will lead to biased estimates. Let $\text{Cov}[\epsilon_1, \epsilon_2] = \rho$. Then, instead we can write

$$\Pr[Y_{cat} = 1, Y_{zip} = 1|\mathbf{x}] = \Phi_2(\beta_1\mathbf{x}, \beta_2\mathbf{x}; \rho) \qquad (8)$$

where $\Phi_2$ is the cumulative distribution function of the bivariate normal with zero means and unit variances.

To capture the association between using Groupon and experimentation we model the latent experimentation functions $Y_{cat}^* = \beta_1\mathbf{x} + \epsilon_1$, and $Y_{zip}^* = \beta_2\mathbf{x} + \epsilon_2$ by incorporating the following effects: Groupon user, Groupon mention, Deal category, Deal city, and log(Review count). These variables have been discussed previously, with the exception of Review count, which we use to control for the fact the longer review histories are more likely to contain a match. As before, the model can be estimated using maximum likelihood. The results of estimating the model are shown in Table VII for completeness, but we interpret the model in terms of the average predicted probabilities (analogous to average marginal effects presented earlier) depicted in Table VIII. As before, we find that Groupon users are on average more likely to engage in experimentation when mentioning (and we assume, using) Groupons. For example, consider the rightmost column of Table VIII, corresponding to the likelihood of a review being one for which there is neither a category nor a ZIP match, i.e. indicating a high degree of experimentation. For non-Groupon users and for Groupon users mentioning Groupons, this likelihood is over 30%, while for Groupon users (when not mentioning Groupon), this likelihood is much lower, less than 10%. Most surprisingly, even after controlling for their longer review histories, Groupon users are less likely on average to experiment than non-Groupon users without coupons, a result we find curious.

## 7. CONCLUSION

We have examined a number of hypotheses for explaining what we have dubbed the Groupon effect. While there remain challenges in trying to exactly quantify the different issues at play, we have shown that a combination of poor business behavior, Groupon user experimentation, and an artificially high baseline all play a role.

Our compilation of evidence about Groupon users, as presented in Tables I and IV, highlights that on average, Groupon users provide detailed reviews that are valued more highly by their Yelp peers. Contrary to what some would believe, the evidence suggests that Groupon users are no more critical than their peers, although they may well be experimenting with a new business when using a Groupon. It seems odd that any business would incur marketing costs only to treat new customers poorly, but as the FTD example shows, sometimes businesses do unusual things, such as treating potential long-term customers as a sunk cost rather than as an opportunity. Although it seems obvious, we would advise businesses to treat Groupon customers as well as (or possibly better than?) their other customers to avoid negative reputational impact.

# Appendix

## Justification for relaxing the parallel regressions assumption

To test the parallel regressions assumption for various key model parameters, we performed a series of likelihood ratio tests on the fully constrained specification by separately relaxing the constraints on coefficient of each explanatory variable. The results are presented in Table IX. In all cases, we observe a significant test statistic, strongly suggesting that the parallel regressions assumptions is violated.

Table IX: Tests of the parallel regressions assumption for various model parameters. Test statistics for a $\chi^2$ distribution are shown.

|                 | $\chi^2$  | $p < \chi^2$ | Degr. of Fr. |
|-----------------|-----------|--------------|--------------|
| Groupon mention | 188.0890  | 0.000        | 3            |
| Groupon user    | 1128.4384 | 0.000        | 3            |
| Groupon period  | 810.1980  | 0.000        | 6            |
| Review rank     | 477.4282  | 0.000        | 3            |

## Additional tables mentioned in the paper.

In our previous paper [Byers et al. 2012], we presented Figure 6 as evidence of the Groupon effect. As we describe in the main text, this figure is much more ambiguous, in large part because the data is presented at a monthly granularity on a smaller dataset. We believe our new data makes the existence of the Groupon effect much clearer.

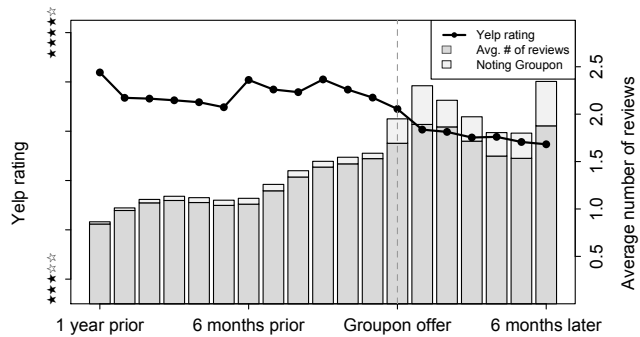

Fig. 6: The average Yelp star-rating for merchants before and after their Groupon offers (line-chart), and the average number of reviews per merchant per month (bar-chart). Reproduced from [Byers et al. 2012].

Table X, which we described in considering the hypothesis that Groupon businesses are more likely to be "bad" businesses, provides the magnitude of decline across categories with at least 100 deals.

Table X: Changes in ratings around the Groupon offer date by category for categories with at least 100 deals.

|  | Change in Rating | |
|---|---|---|
|  | Mean | Std. Dev. |
| Arts and Entertainment | -0.05 | 1.03 |
| Automotive | -0.34 | 1.28 |
| Beauty & Spas | -0.28 | 1.15 |
| Education | -0.16 | 1.02 |
| Food & Drink | -0.25 | 0.87 |
| Health & Fitness | -0.13 | 1.03 |
| Home Services | 0.42 | 1.65 |
| Nightlife | -0.03 | 0.71 |
| Pets | 0.09 | 1.36 |
| Professional Services | -0.39 | 1.28 |
| Restaurants | -0.18 | 0.75 |
| Shopping | -0.23 | 1.17 |
| Travel | -0.13 | 0.88 |